# Laser-assisted coloration of Ti: oxides or nanostructures?

**E.V. Barmina[1], E, Stratakis[2], C. Fotakis[2], and G.A. Shafeev[1]**

[1]Wave Research Center of A.M. Prokhorov General Physics Institute of the Russian Academy of Sciences, 38, Vavilov street, 119991, Moscow, Russian Federation
[2] Institute of Electronic Structure and Laser, Foundation for Research & Technology—Hellas, (IESL-FORTH), P.O. Box 1527, Heraklion 711 10, Greece

## Introduction

The interaction of short pulse laser radiation with conducting solids is accompanied by various instabilities, which may lead to the formation of surface textures. In general, these are divided into large- and small-scale textures. The large-scale textures result from the capillary motion of the melt, produced by laser irradiation of the target. Under ordinary conditions, their characteristic length scale is tens of μm. On the other hand, the size of the finer surface structures is of the order of the laser wavelength, and they originate from the interference of the incident laser radiation and the surface electro-magnetic wave (SEW) generated on the ~~of~~ solid surface. The resulting interference pattern is stationary because the waves are in phase, and the corresponding surface morphology is determined by a variety of processes (including oxidation, sublimation, and others), whose rate is temperature-dependent. Typical morphology of these structures consists of straight periodic ripples, and their period scales with the laser wavelength. Laser ablation of solids in liquids ~~is~~ also leads to surface texturing of the target, however in this case different mechanisms prevail, owing to fact that the liquid adjoining the melted surface undergoes a phase transition. Short laser pulses give rise not only to superheated liquid but also to a transient zone of elevated pressure close to the target, which may bring the surrounding medium to a supercritical state. As a result of the interaction of the pressure wave with the melt layer on the target surface, its morphology is changed. This surface profile is frozen upon cooling before being smeared by the surface tension of the melt. Nanotextures of this type were first observed on silver [1] and then on gold [2], upon short pulse laser ablation of Ag and Au targets in certain liquids. The textures observed were self-organized nanostructures (NS) that look like an array of periodic surface protrusions. Similarly, NS formation on Aluminum and Tantalum targets has been recently reported [3-5]. In the case of Ta the surface protrusions make up quasi-periodic arrays with their average period is being a function of laser parameters, including laser fluence and pulse duration. In case of femtosecond (fs) laser radiation, two types of NS co-exist, namely the surface protrusions are situated on top of periodic ripples [5]. Surface nanotexturing is accompanied by changes in the absorption spectrum of the target material, so that additional bands emerge near the plasmon resonances of metal NS. Furthermore, such NS can advantageously exhibit Surface Enhanced Raman Scattering (SERS) of adsorbed organic molecules with an enhancement factor being as high as $10^5$ [2,6].

In the present communication we report on the physical and chemical properties of Ti ablated in liquids with picoseconds (ps) laser pulses. It is found that the formation of NS on Ti surface under its laser ablation in various liquids is accompanied by its strong visible coloration, and the corresponding color correlates with the NS morphology. The results are discussed in conjunction with previous observations of laser-induced Ti coloration reported by Morenza *et al* [7].

## Experimental

Nanostructuring of a Ti target has been achieved upon its laser ablation in liquids, either water or ethanol, with ps laser sources. The first laser used was a KrF excimer laser emitting at 248 nm with pulse duration of 5 ps. The second one was a tripled Nd:YAG laser emitting 150 ps pulses at the wavelength of 355 nm. Ti plates of 99.9% purity were used as received without any polishing. The morphology of the surface was characterized by Field Emission Scanning Electron Microscopy (FE SEM). The absorption spectra of various Ti surfaces were acquired using a Perkin-Elmer spectrometer Lambda-950 equipped with integrating sphere in the range 250 – 2500 nm. All the spectra were taken with Spectralon™ as a reference reflector. Raman spectra were obtained using a micro-Raman spectrometer (NICOLET ALMEGA XR) with a 473 nm laser as an excitation source.

## Results

### Optical Properties

The laser exposed areas of the target take on eye-visible coloration, with the intensity and color being dependent on the number of laser shots. At low number of pulses the surface looks yellow, then blue, and with further increase of shots it turns into purple. The corresponding absorption spectra are presented in Fig. 1, showing that the sample irradiated in water, looking purple in appearance, absorbs more than 90% of light in green region.

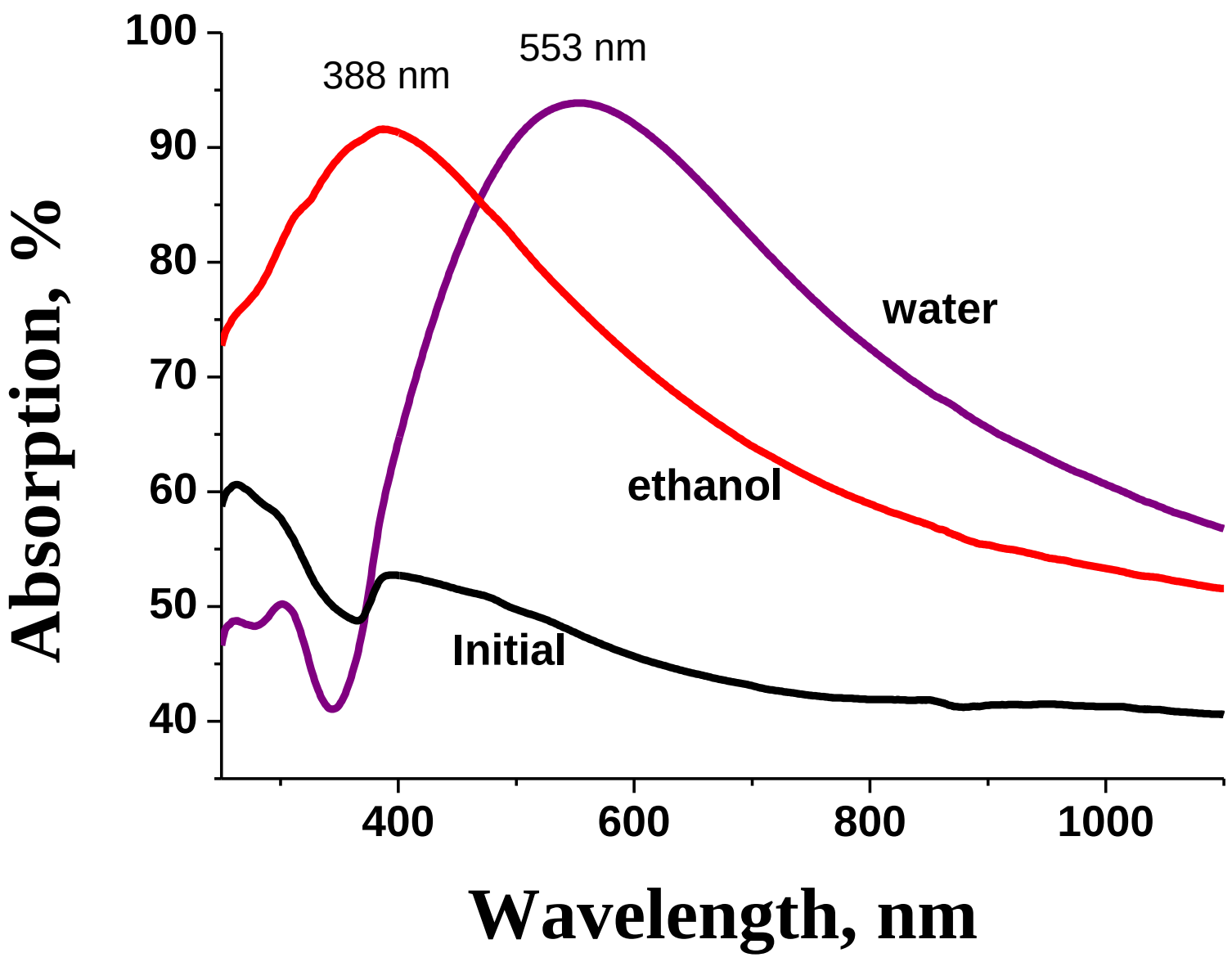

Fig. 1. Absorption spectra of pristine (black) and Ti surfaces exposed to 5 ps laser pulses in ethanol (red) and water (purple).

Similar coloration is also observed under exposure of Ti in water with 150 ps laser pulses. The absorption spectrum of this surface is presented in Fig. 2. The maximum of absorption in this case is situated at around 530 nm though this position is only a matter of the number of laser shots; upon increasing their number the intensity of coloration increases, and its maximum shifts further to the red.

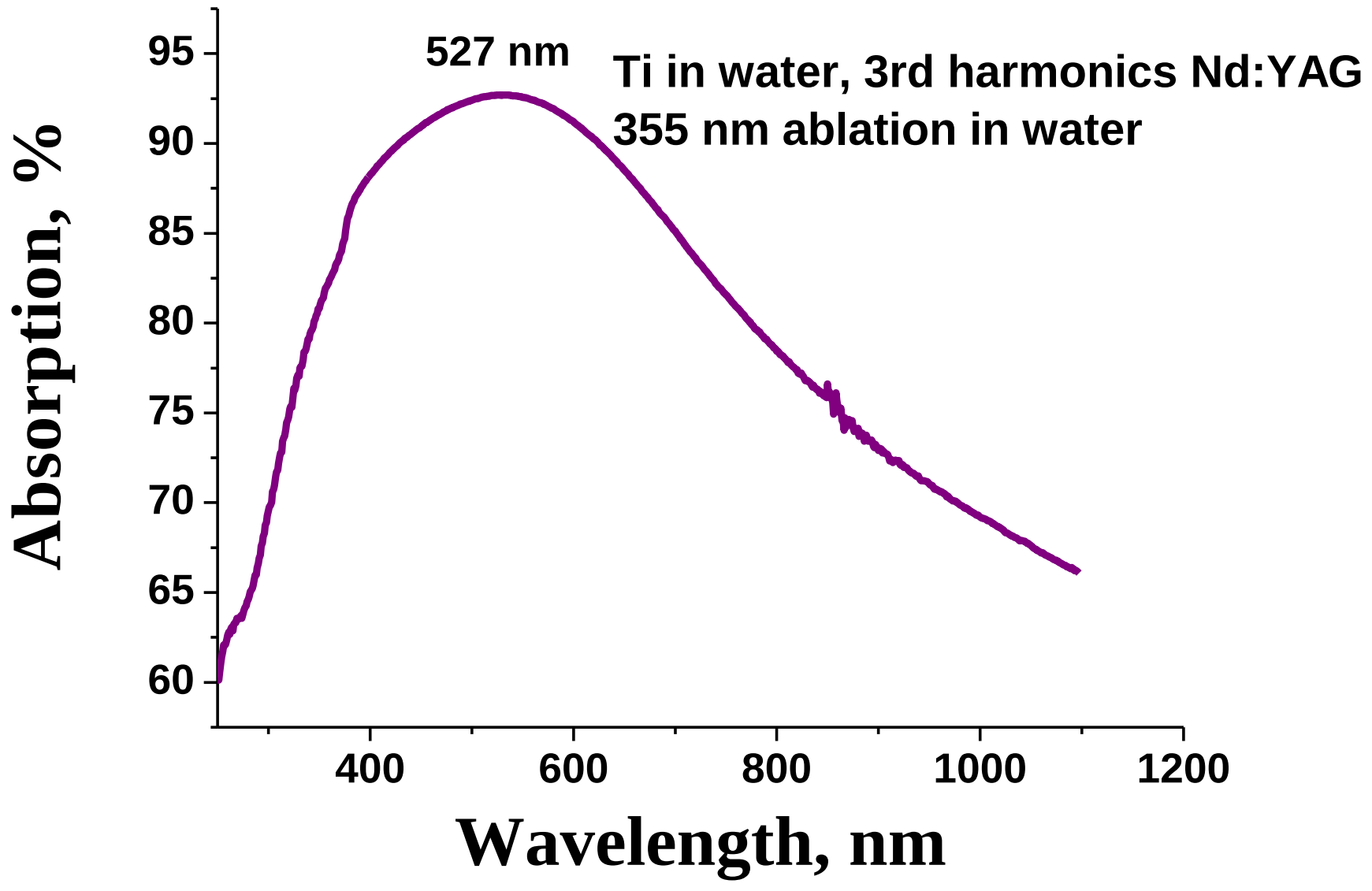

Fig. 2. Absorption spectra of Ti surface exposed to 150 ps laser pulses in water.

**Morphology**

FESEM views of the Ti target exposed to ps laser pulses in liquid media show the presence of NS in laser-exposed areas. Figs. 3a and b show typical images of the Ti surface exposed in ethanol and water respectively. In particular, NS produced in water have a conical shape with a tip diameter of about 100 nm. It should be noted that the irradiation time needed to produce visible coloration is higher in ethanol, compared to water. This is probably due to the proximity of the laser wavelength (248 nm) to the edge of fundamental absorption in ethanol.

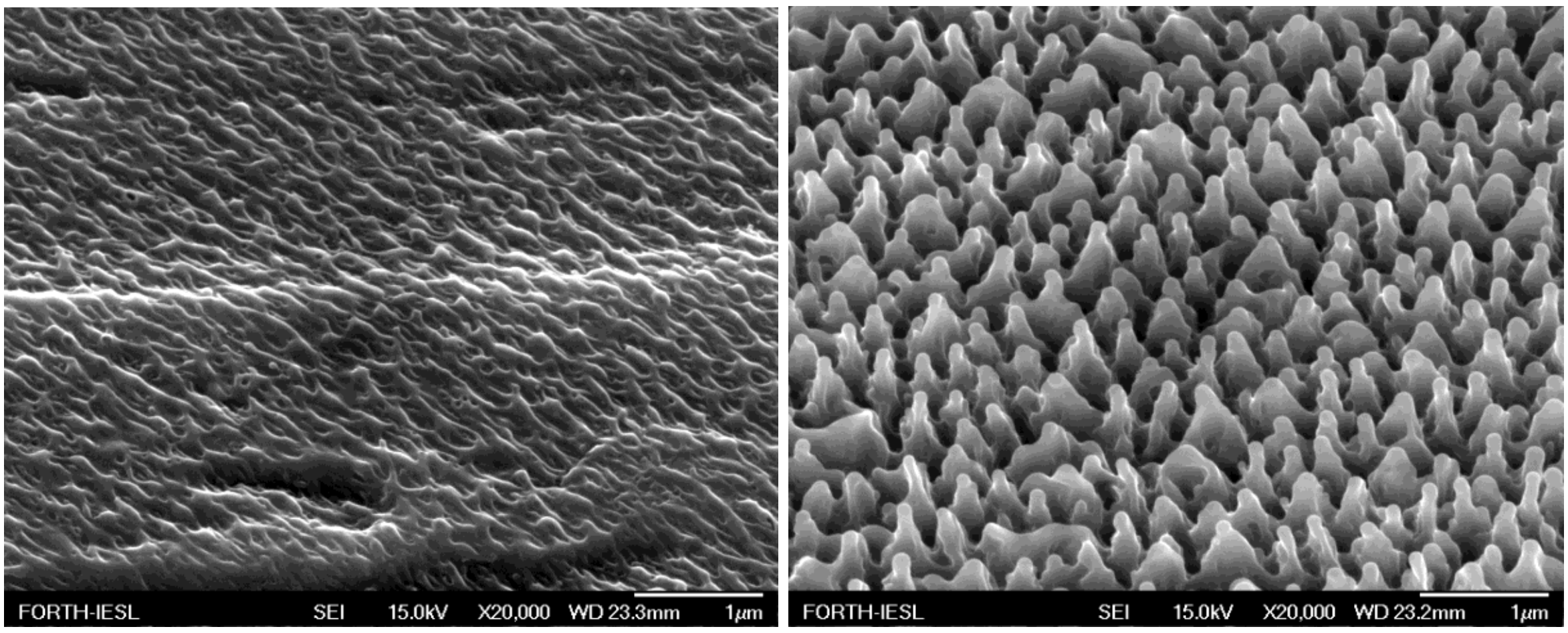

Fig. 3. Morphology of Ti surface exposed to 5 ps laser radiation in ethanol (a) and in water (b). The irradiation is performed with 1200 pulses at a fluence of 1 J/cm2 in ethanol and 600 pulses at a fluence of 0.4 J/cm$^2$ in water.

**Density of nanospikes vs number of laser shots**

Fig. 4 shows FESEM views of a Ti surface ablated in water by different number of pulses at the same fluence. For low numbers of laser shots a planar structure comprising walls of various lengths is developed. The 3D nanostructure formed at high numbers of shots comes as the combination of planar walls with nanospikes situated on top of them. The height of spikes is as about 400-500 nm with lateral dimensions about 100 nm. The width of underlying walls is also of order of 100 nm, while their length amounts to 1 μm.

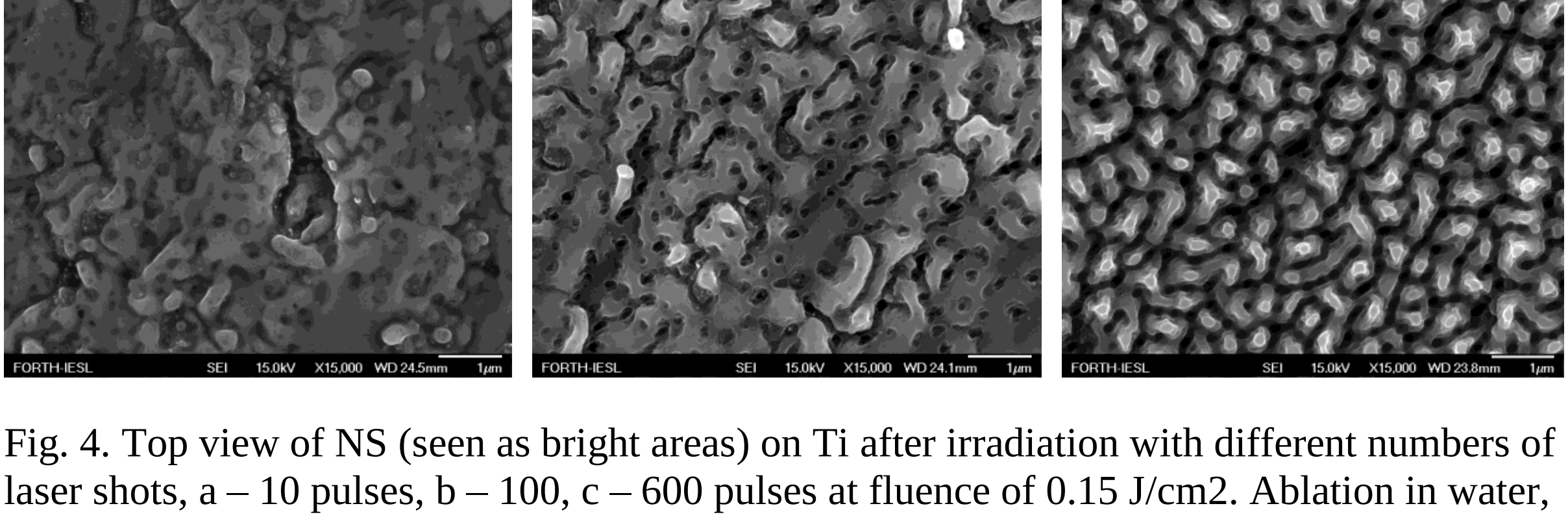

Fig. 4. Top view of NS (seen as bright areas) on Ti after irradiation with different numbers of laser shots, a – 10 pulses, b – 100, c – 600 pulses at fluence of 0.15 J/cm2. Ablation in water, 5 ps, 248 nm. Scale bar denotes 1 μm.

The corresponding plot of the NS density against the number of laser shots at various fluences is presented in Fig. 5. It can be seen that the NS density reaches a plateau value very rapidly at low laser fluence. On the contrary no saturation is observed at the highest fluence studied. This may be due to the fact that such high fluence exceeds by far the ablation threshold of Ti, and the ablation process is accompanied by a high production rate of nanoparticles dispersed in the surrounding liquid.

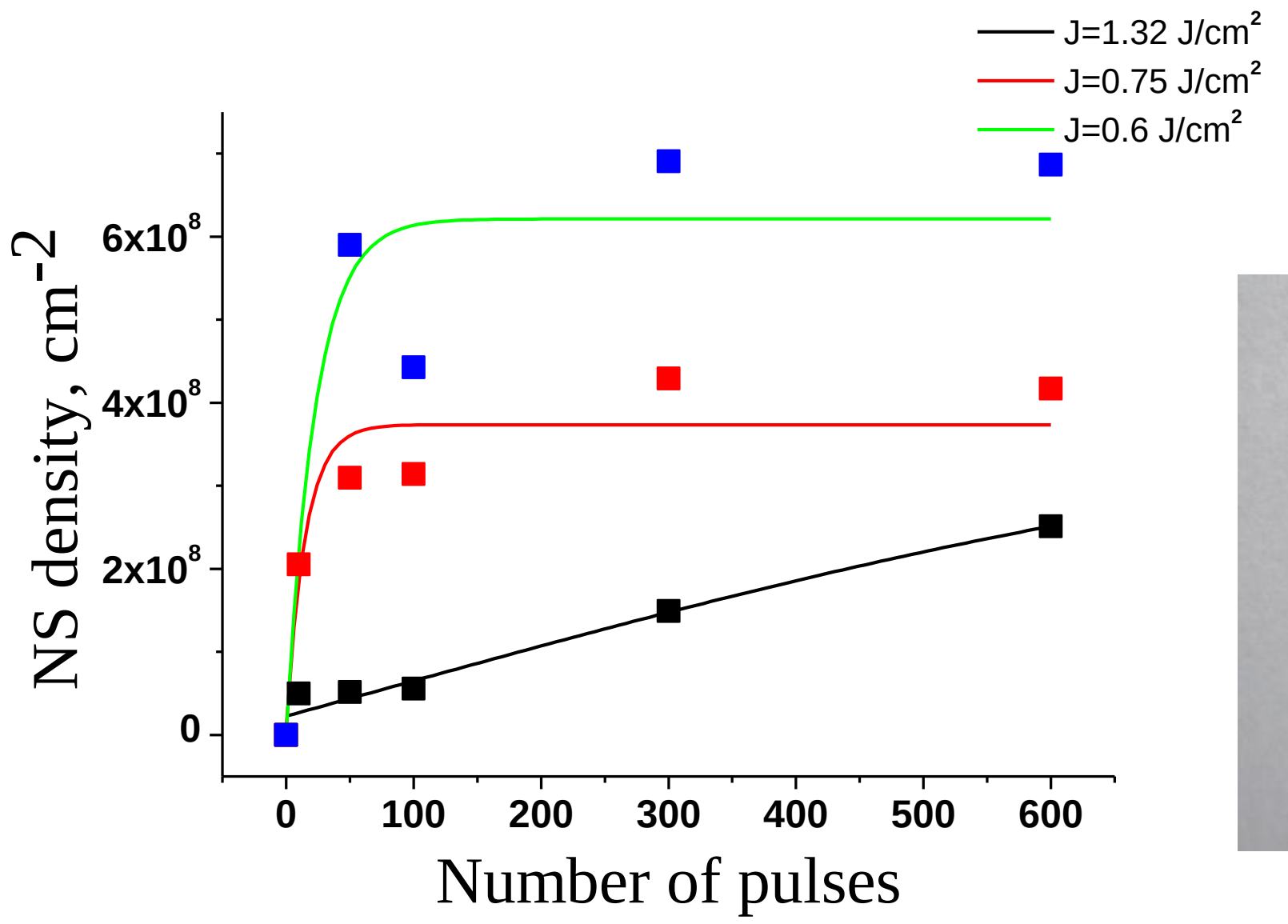

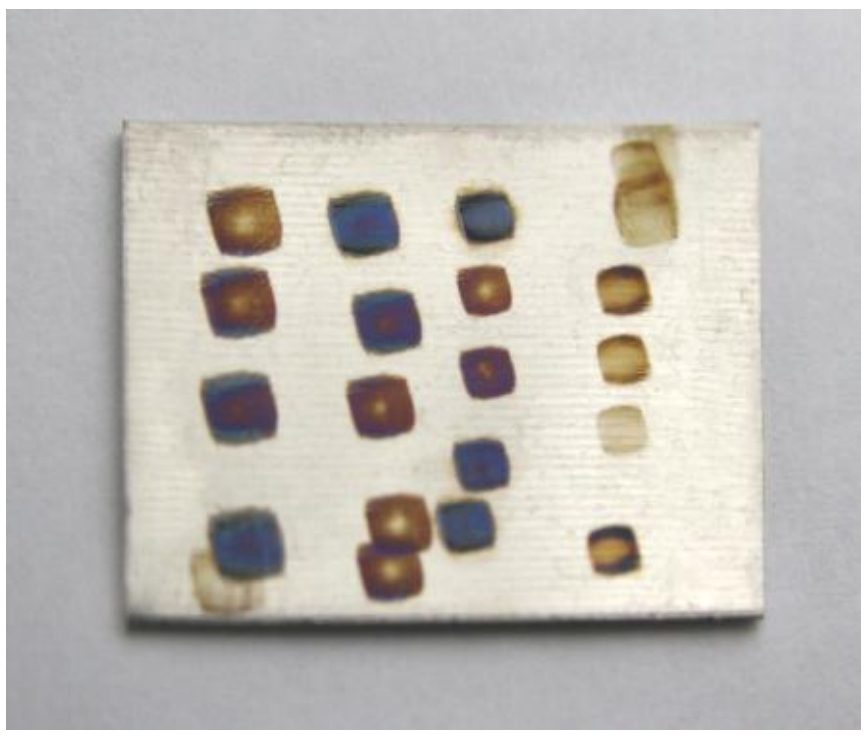

Fig. 5. Density of NS as a function of the number of laser shots for various laser fluences. The ablation process is performed in water with 5 ps laser pulses of 248 nm wavelength (left). Variation of colors at different number of laser shots (right).

**Raman spectroscopy**

Raman spectra of the initial and irradiated surfaces are presented in Fig. 6.

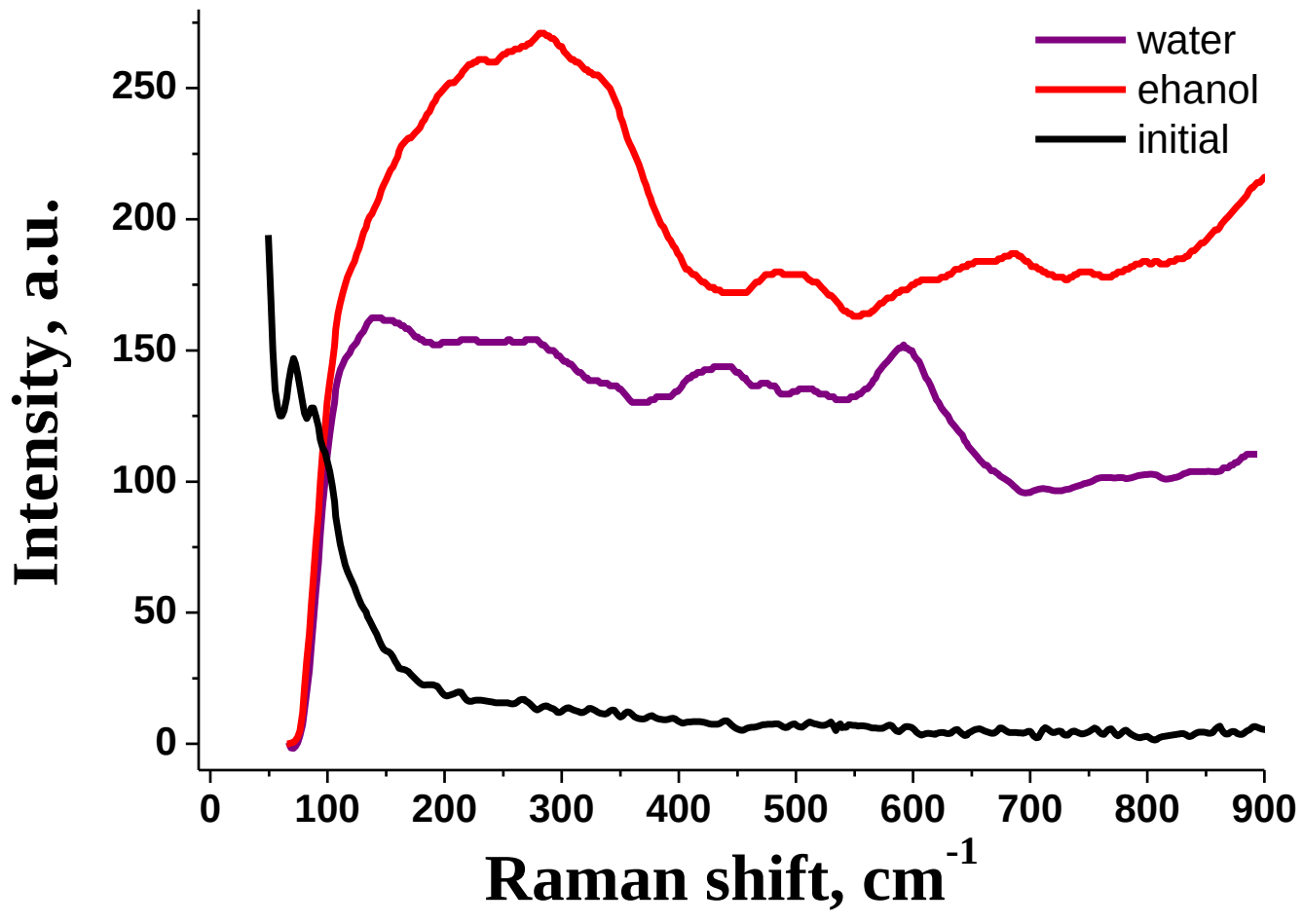

Fig. 6. Raman spectra of of pristine (black) and Ti surfaces exposed to 5 ps laser pulses in ethanol (red) and water (purple).

As expected, the initial Ti surface shows no Raman peaks, while the spectrum of Ti exposed in water is similar to that of rutile $TiO_2$. However, the oxide layer on the structured Ti surface should not be thick, since no charging of the surface was observed during its FESEM characterization. Furthermore, rutile is transparent in the visible and therefore cannot contribute to the coloration of the laser-exposed areas.

**Discussion**

Coloration of Ti by its laser ablation in air has been reported by Morenza et al [7]. In this work, a 300 ns Nd:YAG laser was used, and a typical fluence on the target was between 50 and 300 J/cm$^2$. The authors reported different coloration of the Ti surface ranging from blue to red colors. X-ray characterization revealed the presence the Raman-silent phases TiO and $Ti_2O$ in laser-exposed areas. Raman characterization of the exposed Ti surface showed the combination of rutile, amorphous, and crystalline $Ti_2O_3$ phases. However, the authors pointed out that there is an incomplete correlation between the coloration of the laser-treated areas and their composition determined by Raman analysis. Accordingly, red and blue-colored areas exhibited virtually the same Raman spectrum. The authors reached the reasonable conclusion that some structural features of the process might be responsible for the observed coloration of the laser treated areas.

The coloration of Ti ablated in liquids with ps laser pulses can be attributed to the formation of NS on it. Similar to previous observations on coloration of metal surfaces decorated with NS, such coloration can be attributed to the plasmon oscillation of electrons on the NS formed. Coloration due to surface oxidation may be ruled out, as the laser fluence required for the formation of NS is two orders of magnitude lower than that used for laser-induced oxidation of Ti in air. The same is true for the pulse duration, which is four orders of magnitude lower than used in [7]. It appears that, ablation in liquid environment inhibits the oxidation of the target due to lower oxygen content. The above indicate that the oxide thickness on laser-processed areas is of order of the thickness of the native oxide and does not exceed few nanometers. This consideration is supported by X-ray Photoelecron Spectroscopy (XPS) analysis of both pristine Ti surface and laser-colored areas presented in Fig. 7. XPS spectra show the presence of the peaks of both metallic Ti and $TiO_2$, similar to that of pristine Ti. Since the penetration depth of electrons that excite X-ray emission is about 10 nm at the energy used, one may conclude that the thickness of oxide layer in laser-exposed areas is at least of this order.

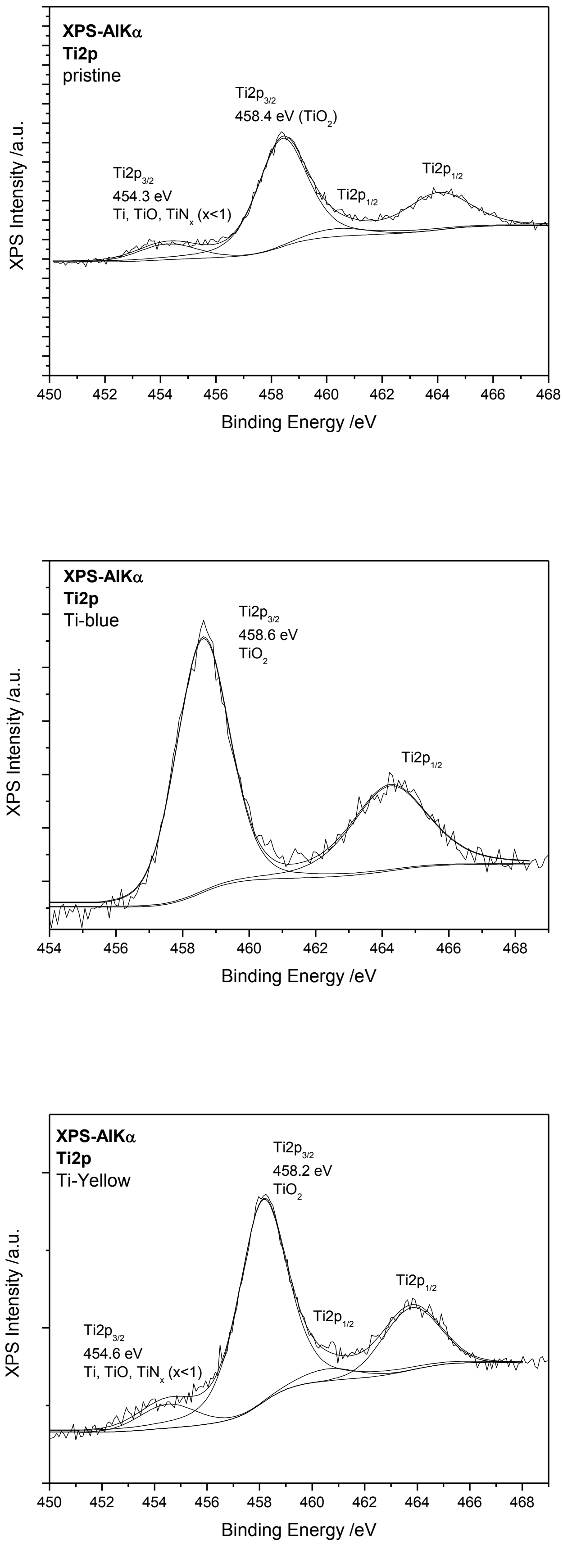

Fig. 7. XPS data on the binding energy of the Ti 2p peak in pristine (a), blue (b), and yellow areas (c) of the Ti target.

The different absorption spectra shown by the ablated Ti samples correspond to differences in the NS morphology. Indeed, yellowish coloration observed on samples ablated in ethanol corresponds to small nanowalls and blue/purple coloration corresponds to elongated nanospikes with aspect ratio 1:4 – 1:5 (see Fig. 3). In general, elongated nanoparticles are characterized by two plasmonic bands, one of which is so called transverse plasmon resonance and the second is the longitudinal one. The position of the latter depends on the aspect ratio of the nanoparticle, the higher is the ratio the more red-shifted is its position. The same is true for the walls which are adjacent to the target surface. The first plasmon resonance corresponds to oscillations of electrons across the nanowall while the second one to oscillations along it. So the absorption spectra of NS on Ti shown above are the mixture of both transverse and longitudinal plasmon resonances in both walls and spikes.

The attribution of Ti coloration to the NS formation implies proximity of the absorption spectrum of NS to that of Ti nanoparticles. It should be noted here that the plasmon resonance peak of spherical Ti nanoparticles moves from the UV region to the visible as soon as their size reaches 50-100 nm [9]. Fig. 8 presents the theoretical plasmon absorption spectrum of larger Ti NPs in comparison to normalized experimental absorption data of the nanostructured Ti surface. Best fit of the experimental absorption curve is achieved assuming spherical Ti nanoparticles of 100 nm in diameter embedded into a medium with refractive index of 1.42. Note that the value of 100 nm is close to the average size of nanospike tips present on the structured Ti surface (see Fig. 3b). A refractive index exceeding 1 should be due to the oxide layer that is formed on the NS surface. According to Raman spectra this oxide layer is rutile, which exhibits a refractive index equal to 2.6. The discrepancy between the oxide index indicated by the experiment and that suggested by theory is presumably due to the very small thickness of the oxide layer, so that the effective refractive index lies between 1 and 2.6. This is further supported by the presence of the peaks of metallic Ti in all XPS spectra in Fig. 7 indicating that the thickness of the oxide layer on laser-exposed areas is of the same order of magnitude as on non-exposed areas. The mismatch of the corresponding spectra in the red region should be ascribed to the non-spherical shape of NS. As pointed out above the spectrum of such NS is composed of two different plasmonic peaks, the one corresponding to transverse and the second one to longitudinal oscillations (relative to the particle long axis). The longitudinal peak should be red-shifted with respect to the transverse, which is in qualitative agreement with the red tail of absorption observed in the experimental curve of Fig. 8.

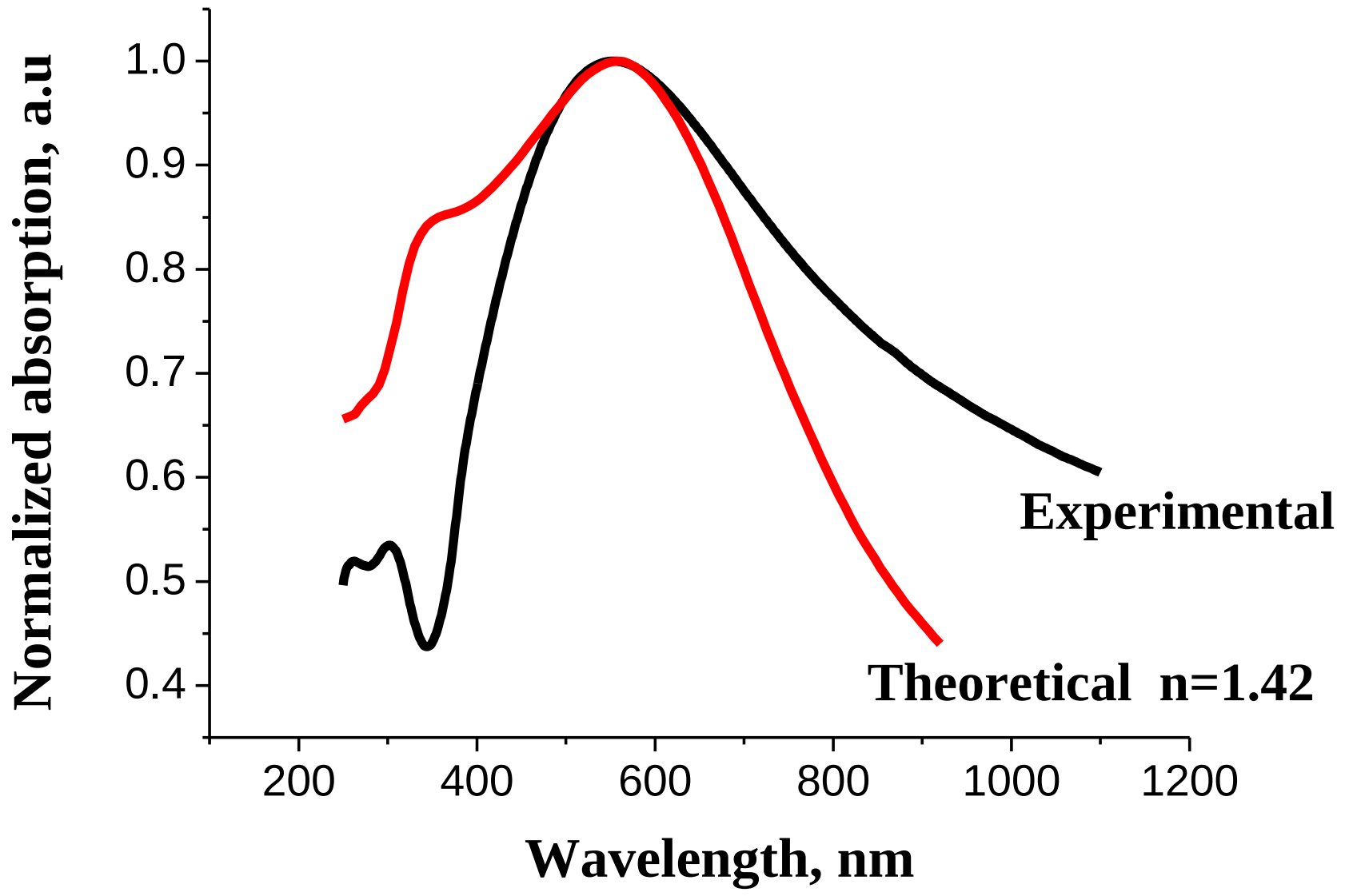

Fig. 8. Fitting the experimental absorption data of nanostructured Ti surface with the theoretical plasmon resonance curve of spherical Ti nanoparticles of 100 nm in diameter.

It is likely that nanostructures on Ti formed under its ablation in air with 300 ns laser [7] might contribute to its coloration. Ti easily evaporates under laser heating, especially at the elevated laser fluence of 300 J/cm$^2$ used in that work. Therefore, the recoil pressure of its vapors may be responsible for the nanospike formation. The spatial resolution of the images of the Ti surface presented in [7] is not sufficient to conclude for their presence. Furthermore, in the case of 300 ns laser pulses, the oxide layer thickness grown on the surface is sufficient to observe well-distinguished X-ray diffraction peaks. On the other hand, exposure of Ti to ps pulses in liquids leads to thinner oxide, and thus the coloration of the exposed areas should be mostly due to the formation of surface NS.

Coloration of metallic target (Al) exposed to fs laser radiation has been reported by Guo et al [8]. It is observed that after ablation at sufficiently high laser fluence, the target takes on a gold-yellow tint. Later, this coloration has been attributed to the formation of NS on Al surface under its ablation with fs laser pulses either in air or in liquid environment [3]. The mechanism of NS formation suggested in Refs [1] and [2] is based on the instability of the melt surface patterned by the recoil pressure of the surrounding liquid vapors. In case of ablation in air the recoil pressure is due to the evaporated target material. This is the reason for the observation that the NS formation in liquid environment requires near-threshold laser fluence, while in air the fluence should be much higher to provide sufficient magnitude to the recoil pressure of the evaporated material.

It is important to note that the colors observed on Ti target after its laser ablation in liquids can also be observed when a Ti substrate is sputtered with Ti nano-entities in vacuum using fs laser radiation. Under the laser exposure the Ti target evaporates, and a Ti substrate placed nearby but not exposed to laser radiation is covered by the sputtered material. The resulting surface comprises of Ti nanoclusters of size between 30 and 50 nm (Fig. 8) and the covered areas show all the color variations observed on the samples structured by ps laser radiation in liquids, i.e. from yellowish to deep purple. The corresponding absorption spectrum, shown in Fig. 9, is similar to that of the surface ablated in water, while the peak position indicates only the average size of Ti clusters present on the surface. It is concluded that the coloration observed on this sputtered sample is due to the presence of nanoclusters on its surface.

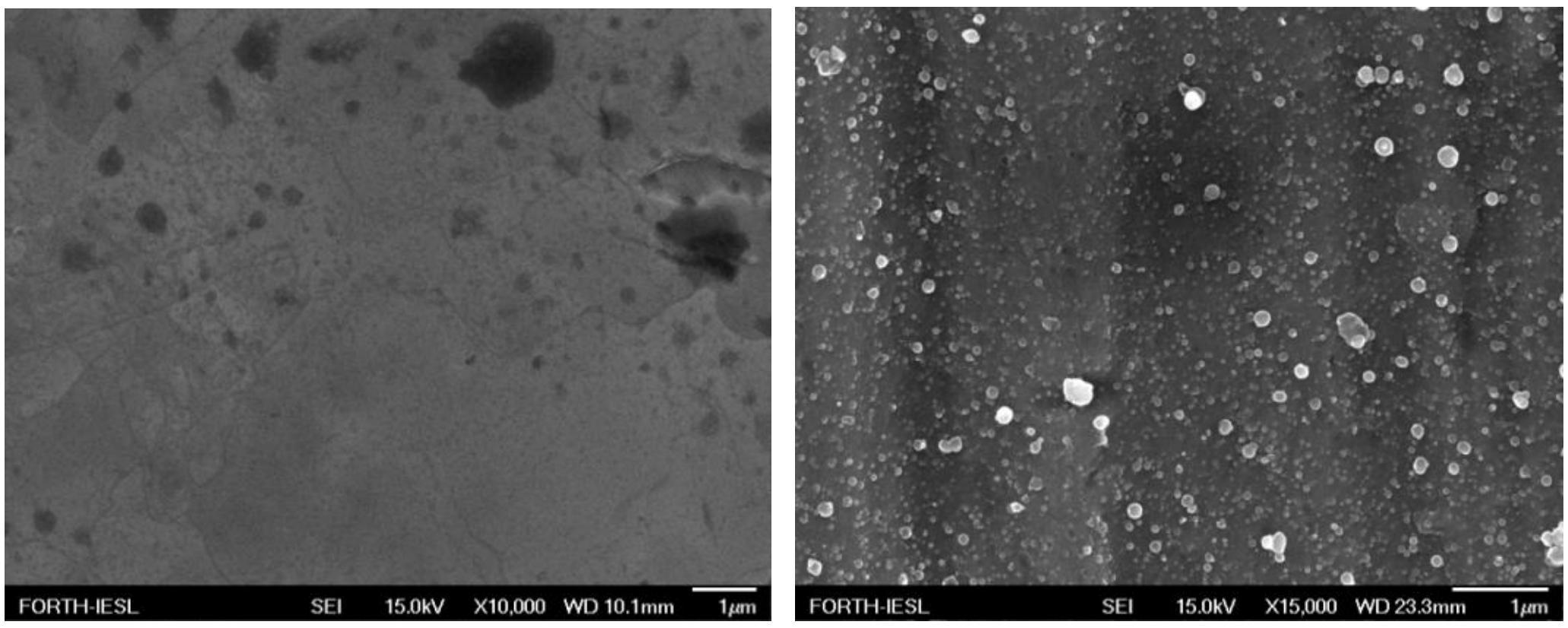

Fig. 8. FE SEM view of the pristine Ti surface (left) and the surface covered by sputtered material from Ti target in vacuum under laser exposure (right). Scale bar denotes 1 μm

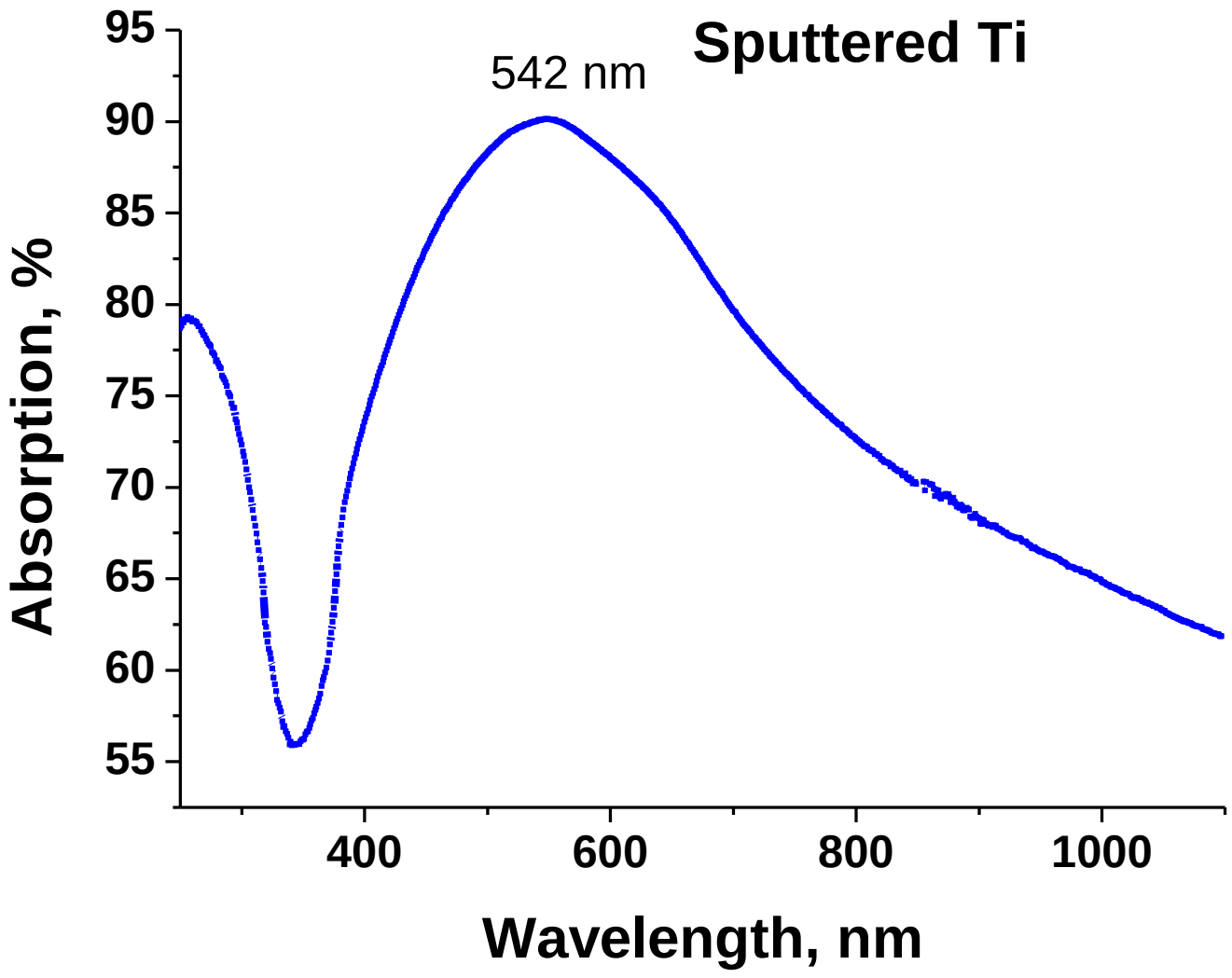

Fig. 9. Absorption spectrum of a Ti plate covered by the sputtered material ejected from a Ti target in vacuum.

Raman spectra of Ti sputtered in vacuum on a Ti substrate are similar to those obtained under laser exposure of Ti in water. The spectra corresponding to two regions with different visible coloration (red and blue) are presented in Fig. 10. The common bands in the vicinity of 140 and 620 cm$^{-1}$ can be attributed to rutile. Note that sputtered Ti clusters that might be formed under its ablation in air with 300 ns laser pulses [7] can hardly contribute to the sample coloration. Indeed, clusters evaporated in air are rapidly oxidized and are deposited on the substrate only like stoichiometric oxides. The Ti surface exposed in vacuum with fs pulses shows no coloration at all. This is coherent with recent observations on the formation of periodic gratings on Ti exposed either in air or in water to fs laser radiation [10].

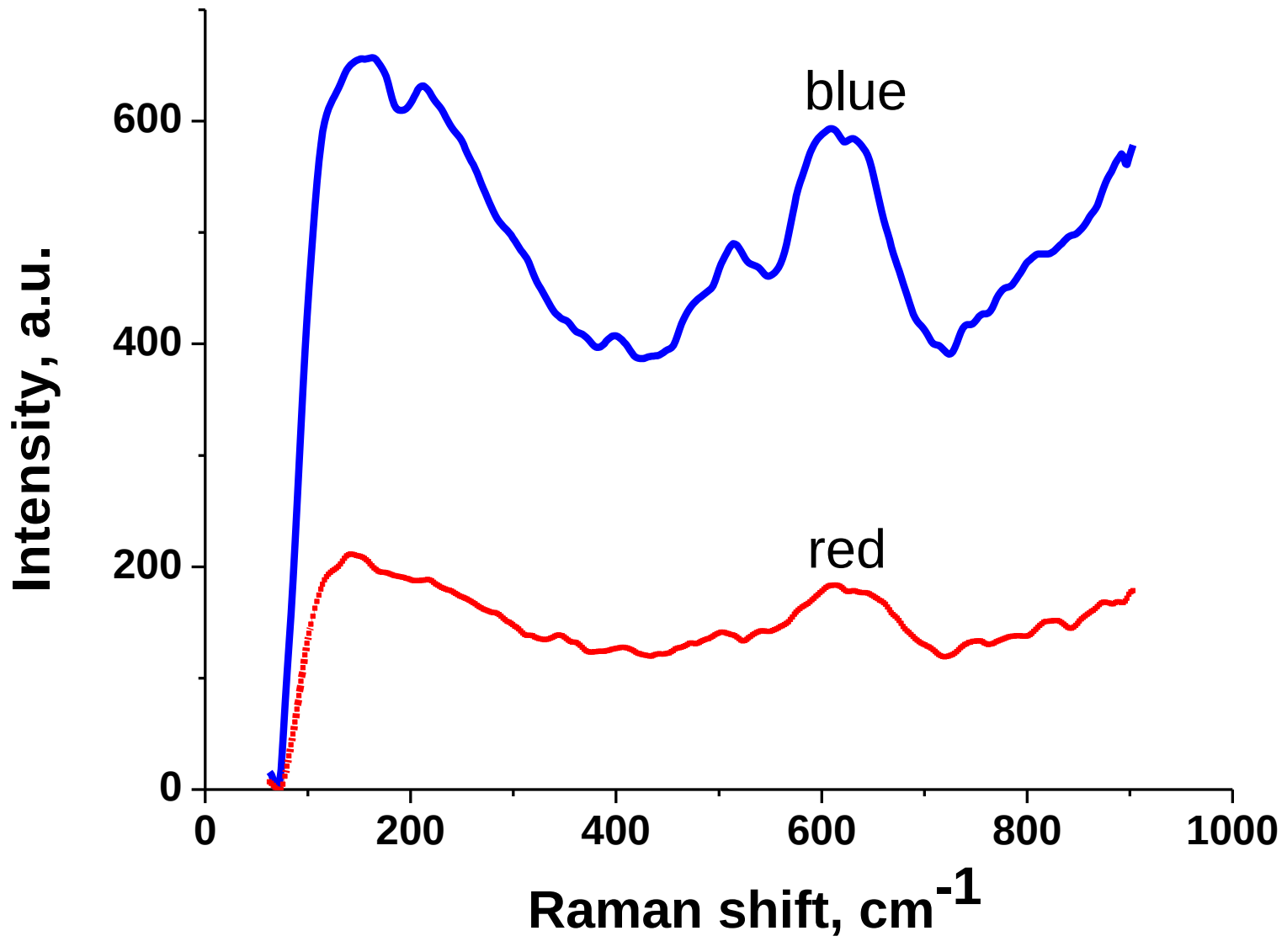

Fig. 9. Raman spectra of Ti clusters sputtered on a Ti substrate in vacuum under ablation with a femtosecond laser. Labels indicate the visible coloration of the samples areas.

One more argument in favor of coloration via formation of NS on Ti rather than its oxidation is the value of the absorption coefficient of the laser-exposed Ti surface. If the absorption near 550 nm (Fig. 1) is due to some absorbing (oxide) film, then it can be estimated. Let us suppose that the thickness of this film is 100 nm, which is merely above the real value since no charging of the Ti surface is observed under FE SEM characterization. Calculations show that the absorption coefficient of this hypothetical film should be as high as $1.15\times10^{5}$ cm$^{-1}$. A thinner film should possess even higher absorption coefficient to account for the absorption of the Ti surface in the visible. This high value of absorption coefficient is typical for electronic absorption having high oscillator force. On the contrary, Ti oxides are virtually transparent in the visible.

## Auto-SERS effect

One may suggest that the Raman spectra observed on a Ti surface decorated with NS correspond to oxides of various compositions. However, the oxide thickness should be close to that of the native oxide of the pristine Ti surface. The relatively high intensity of oxides peaks in Raman spectrum (compared to pristine Ti surface that is also oxidized) may be due to the enhancement of the scattering effect by the underlying Ti NS. Therefore, the capability to observe the oxide phases in Raman spectra is firstly due to the higher specific area of the nanostructured Ti surface and secondly, due to possible enhancement of the Raman signal by the underlying Ti NS.